\documentclass[accepted]{article}

\usepackage{microtype}
\usepackage{graphicx}
\usepackage{subcaption}
\usepackage{booktabs} 

\usepackage{hyperref}

\usepackage{icml2026}

\usepackage{amsmath}
\usepackage{amssymb}
\usepackage{mathtools}
\usepackage{amsthm}
\usepackage{enumitem}

\usepackage[capitalize,noabbrev]{cleveref}

\theoremstyle{plain}

\theoremstyle{definition}

\theoremstyle{remark}

\usepackage[textsize=tiny]{todonotes}

\icmltitlerunning{Learning Interface Breakup: A Geometry-Conditioned Latent Surrogate for Spray Formation}

\begin{document}

\twocolumn[
  \icmltitle{Learning Interface Breakup: A Geometry-Conditioned \\ Latent Surrogate for Spray Formation}
  
  \icmlsetsymbol{equal}{*}

  \begin{icmlauthorlist}
    \icmlauthor{Julius H Ramlau}{cheme,cmac}
    \icmlauthor{Friedrich Hastedt}{cheme,sarg}
    \icmlauthor{Tolga Birdal}{comp}
    \icmlauthor{Ehecatl-Antonio del Río Chanona}{cheme,sarg}
    \icmlauthor{Nausheen S Basha}{cheme,manch}
    \icmlauthor{Omar K Matar}{cheme}
  \end{icmlauthorlist}

  \icmlaffiliation{cheme}{Department of Chemical Engineering, Imperial College London, London, United Kingdom}
  \icmlaffiliation{cmac}{CMAC, CEDAR, an EPSRC Center for Doctoral Training in Cyber-Physical Systems for Medicines Development and Manufacturing, Imperial College London, London, United Kingdom}
  \icmlaffiliation{comp}{Department of Computing, Imperial College London, London, United Kingdom}
  \icmlaffiliation{manch}{Department of Chemical Engineering, The University of Manchester, Manchester, United Kingdom}
  \icmlaffiliation{sarg}{Sargent Centre for Process Systems Engineering, Imperial College London, London, United Kingdom}

  \icmlcorrespondingauthor{Julius H Ramlau}{julius.ramlau25@imperial.ac.uk}

  \icmlkeywords{Machine Learning, ICML, Computational Fluid Dynamics, Adaptive Mesh Refinement, Two-Phase Flow, Transient Flow, Complex Flow, Flow, Interfacial Dynamics, Interface, Atomization, Atomization, Spray Formation, Surrogate Modeling, Interface Breakup, Droplets, Pharmaceutical Manufacturing, Neural Networks, Deep Learning, AI4Physics, Fluid Mechanics, Fluid Dynamics, Artificial Intelligence}
  \vskip 0.3in
]



\printAffiliationsAndNotice{}  

\begin{abstract}
Designing spray nozzles requires predicting how geometry shapes transient two-phase breakup, but high-fidelity volume-of-fluid (VOF) simulations with adaptive mesh refinement (AMR) are too expensive for iterative design exploration. Standard surrogate models are also challenged by this setting because both the liquid--gas interface and the underlying adaptive discretization evolve across time and geometries. We introduce a geometry-conditioned latent surrogate trained on 797 two-phase nozzle simulations that addresses this by encoding the AMR cell-density field, rather than the full multi-channel flow state, as a compact proxy for where the solver concentrates resolution. From this representation, the model reconstructs transient density evolution and nozzle geometry, and a lightweight second stage recovers the remaining flow variables. On held-out simulations, the method accurately captures key interface dynamics while reducing inference time to 0.045\,s per trajectory, corresponding to a speed-up of more than $6\times10^4$ relative to \verb|Basilisk| CFD. These results suggest that AMR refinement structure can serve as a compact and learnable representation for geometry-conditioned surrogate modeling of transient two-phase flows.
\end{abstract}

\section{Introduction}
\label{sec:introduction}

Spray nozzles are central to processes such as spray drying, fuel injection, material deposition, drug delivery, and pharmaceutical manufacturing~\cite{Henriques2022, traverso2023}. In all of these, the measurable results, such as droplet size distribution, dispersion pattern, and mass flux, are determined by a sequence of processes that begins with the nozzle geometry: for fixed fluid properties and operating conditions, the shape determines the transient flow structure, which governs interfacial instability and breakup, and thus the resulting droplet statistics~\cite{villermaux2007}. Small perturbations in geometry can substantially alter these outcomes~\cite{basha2025, wu2025}, yet inverse design over them requires hundreds to thousands of forward simulations. Each forward solution is expensive because resolving thin, rapidly evolving interfaces requires adaptive mesh refinement (AMR), which introduces per-timestep mesh refinement, error estimation, and load rebalancing on top of the already stiff coupling between the Navier--Stokes equations, VOF advection, and surface tension, all of which necessitate small timesteps. As a result, computational cost scales poorly with iterative design exploration~\cite{kulkarni2025}.

\begin{figure*}[tb]
\centering
\includegraphics[width=\linewidth]{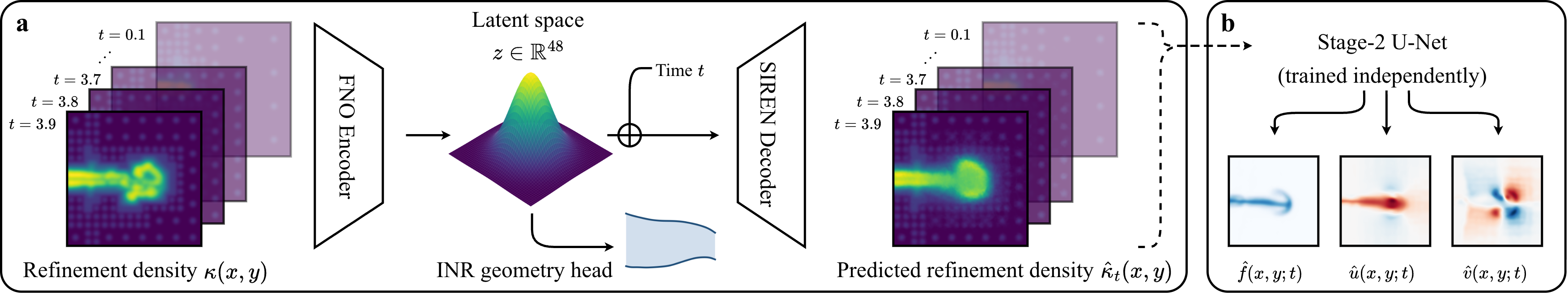}
\caption{\textbf{Geometry-conditioned latent surrogate architecture.}
\emph{(a)} The FNO encoder (four spectral-convolution blocks followed by global average pooling) maps a density snapshot to a time-independent latent $z \in \mathbb{R}^{48}$. The time-conditioned SIREN decoder reconstructs $\hat{\kappa}(x, y; t)$ from $(z, t, x, y)$ via FiLM conditioning with multi-scale spatial Fourier features ($\sigma_s \in \{1, 2, 5, 10, 20, 50, 100\}$) and sinusoidal time features. The geometry head is a 1D implicit neural representation (INR) decoding $(y_\text{upper}(x), y_\text{lower}(x))$ from the same $\mu$. \emph{(b)} The Stage-2 U-Net recovers velocity $(u, v)$ and volume fraction $f$ from the reconstructed density, trained independently.}
\label{fig:architecture}
\end{figure*}

Neural surrogates have emerged as efficient alternatives to costly high-fidelity simulation in fluid dynamics. Operator-learning methods such as Fourier Neural Operators (FNOs)~\cite{li2021} and DeepONet~\cite{lu2021} learn mappings between infinite-dimensional function spaces, while autoencoder-based approaches construct low-dimensional latent representations of flow fields for reduced-order modeling~\cite{linot2023, pan2023, vinograd2025rb, vinograd2026}. Within this landscape, two largely distinct research directions have emerged. First, geometry-conditioned models learn mappings between domain geometry and flow responses, typically in steady or canonical single-phase settings with fixed discretizations~\cite{you2026, Masood2026}. Second, multiphase and interfacial flow surrogates focus on representing highly nonlinear interface dynamics, where discontinuities, topology changes, and adaptive meshes introduce additional challenges for learning~\cite{poels2024, hassan2023, tao2024, cutforth2026}. In the latter case, accurately capturing the geometry of evolving interfaces is itself a central difficulty, and existing approaches primarily focus on faithful state representation rather than explicit geometric conditioning. Despite these advances, the intersection of these two directions remains underexplored. Existing approaches typically either condition on geometry in simplified flow settings~\cite{you2026} or learn latent representations of complex interfacial dynamics without explicit geometric conditioning~\cite{cutforth2026}. As a result, joint representations that simultaneously capture geometry and transient multiphase flow evolution are largely absent from the literature. Such representations would be a natural building block for geometry-conditioned modeling, and inverse design in multiphase systems, where geometry and dynamics are intrinsically coupled.

We take the first step in this direction on a dataset of 797 simulated two-phase nozzle flows. Our key modeling choice is to encode a single, physics-informed channel, the AMR cell-density field, rather than the full multi-channel state. An FNO encoder maps this channel to a time-independent latent; a time-conditioned SIREN~\cite{sitzmann2020} decoder reconstructs the transient density evolution; a geometry head recovers the nozzle walls from the same latent; and a lightweight U-Net~\cite{ronneberger2015} maps the reconstructed density to velocity and volume fraction, completing the flow state (see \cref{fig:architecture}). The pipeline is one-shot (no autoregressive rollout), trained end-to-end on density and nozzle-wall supervision alone, and runs at inference in $\sim$0.045 seconds per trajectory versus 0.75 to 1.5 CPU-hours (12 cores) for the reference \verb|Basilisk| solve.

We make the following contributions: 

\begin{itemize}[leftmargin=*,nosep]
\item We introduce the AMR cell-density field as a solver-informed representation for learning transient two-phase nozzle flows, turning adaptive discretization from a by-product into a useful learning signal.
\item We propose a shared latent representation that couples nozzle geometry and transient flow evolution, enabling both geometric reconstruction and time-dependent flow decoding from the same compact code, including recovery of a usable latent from geometry alone via 
constrained latent-space optimization against the geometry decoder.
\item We show on 797 simulated nozzle flows that this representation supports fast one-shot surrogate inference, reducing trajectory prediction from CPU-hours to milliseconds while preserving key interface dynamics.
\end{itemize}
\section{Methods}

\subsection{Simulation Setup}
\label{sec:simulation}

We generate training data from direct numerical simulations of a pulsed liquid jet undergoing primary atomization in a confined nozzle geometry defined by paired NURBS walls. The flow is modeled as an incompressible, immiscible two-phase system in the one-fluid formulation of the Navier--Stokes equations.

All variables and parameters are reported in nondimensional form. We use the liquid density $\rho_l^\ast$, mean inlet velocity $u_0^\ast$, and inlet diameter $D^\ast = 2R^\ast$ as reference scales for density, velocity, and length, respectively. Time is scaled by $D^\ast/u_0^\ast$, pressure by $\rho_l^\ast {u_0^\ast}^2$, and viscosity by $\rho_l^\ast u_0^\ast D^\ast$. Under this nondimensionalization, the governing equations become
\begin{align}
\nabla \cdot \mathbf{u} =& \ 0, \\
\rho \left( \frac{\partial \mathbf{u}}{\partial t} + \mathbf{u}\cdot\nabla \mathbf{u} \right)
= &  -\nabla p
+ \frac{1}{\mathrm{Re}} \nabla \cdot \left[ \mu \left( \nabla \mathbf{u} + \nabla \mathbf{u}^{\top} \right) \right] \nonumber \\ & + \frac{1}{\mathrm{We}} \mathbf{F}_\sigma, \\
\frac{\partial f}{\partial t} + \mathbf{u}\cdot\nabla f =& \  0,
\end{align}
where $\mathbf{u}$, $p$, $t$, and $f$ denote the nondimensional velocity, pressure, time, and liquid volume fraction, respectively, and $\mathbf{F}_\sigma = \sigma H \mathbf{n} \delta_s$ denotes the surface tension force as the product of surface tension $\sigma$, interfacial curvature $H$, outward-pointing normal $\mathbf{n}$, and Dirac delta function $\delta_s$. The Reynolds and Weber numbers are therefore
\begin{equation}
\mathrm{Re} = \frac{\rho_l^\ast u_0^\ast D^\ast}{\mu_l^\ast},
\qquad
\mathrm{We} = \frac{\rho_l^\ast {u_0^\ast}^2 D^\ast}{\sigma^\ast},
\end{equation}
given the dimensional liquid dynamic viscosity $\mu_l^\ast$ and dimensional surface tension coefficient $\sigma^\ast$.  

We interpolate the local density and viscosity from the phase indicator as
\begin{equation}
\rho = f\rho_l + (1-f)\rho_g, \qquad
\mu = f\mu_l + (1-f)\mu_g,
\end{equation}
where $(\rho_l,\mu_l)$ and $(\rho_g,\mu_g)$ denote the nondimensional liquid and gas properties.

The 
liquid phase entering the nozzle is initialized as a cylindrical column with
nondimensional radius $R = 1/12$ and initial length $L = 0.025$ in solver coordinates, corresponding to an inlet diameter of $D = 2R = 1/6$. We set $\rho_l = 1$ and $\rho_g = \rho_l/27.84$, yielding a density ratio $\rho_l/\rho_g = 27.84$. While this value is lower than that of many physical liquid–gas systems (e.g. water–air, $\mathcal{O}(10^3)$), it is a deliberate and widely adopted compromise in direct numerical simulations (DNS) of primary atomization. High density ratios introduce strong discontinuities in material properties across the interface, which significantly increase the stiffness of the pressure–velocity coupling in VOF formulations and can lead to numerical instabilities and solver divergence, particularly in large parametric studies. This consideration is especially relevant here, where a total of 797 simulations are performed across a broad range of nozzle geometries, and robustness of the solver is critical.

Moderate density ratios such as $\mathcal{O}(10^1$–$10^2)$ are therefore commonly employed to ensure numerical tractability while retaining the essential physics of primary atomization. In particular, the qualitative dynamics governing Kelvin–Helmholtz instability growth, interfacial wave development, ligament formation, and breakup topology are preserved, even if quantitative droplet statistics may differ from high-density-ratio counterparts. Similar choices have been reported in recent DNS studies of atomizing jets (e.g.~\cite{kulkarni2025}), and should be interpreted as a standard trade-off between physical fidelity and computational robustness.

The liquid viscosity is chosen such that the liquid-phase Reynolds number is $\mathrm{Re} = 10^3$, and the gas viscosity is scaled proportionally, $\mu_g = \mu_l/27.84$, so that the viscosity ratio matches the density ratio. This choice yields equal kinematic viscosities in the two phases, which further alleviates numerical stiffness at the interface and improves stability of the coupled solver without altering the dominant inertial and capillary mechanisms governing the flow.

To trigger interfacial instabilities and subsequent breakup, we impose a time-periodic inlet velocity,
\begin{equation}
u_{\mathrm{in}}(t) = u_0 + a_u \sin\left( \frac{2\pi t}{T_0} \right),
\end{equation}
with nondimensional mean speed $u_0 = 1$, perturbation amplitude $a_u = 0.05$, and period $T_0 = 0.1$. Surface tension is set to $\sigma = 3 \times 10^{-5}$ in solver units, which under the above scaling corresponds to a Weber number of
\begin{equation}
\mathrm{We}
=
\frac{\rho_l u_0^2 D}{\sigma}
=
\frac{1 \cdot 1^2 \cdot (1/6)}{3 \times 10^{-5}}
\approx 5.56 \times 10^3.
\end{equation}
This parameter regime produces transient interface dynamics characterized by ligament formation and droplet detachment.

Simulations are performed with adaptive mesh refinement up to level 9. Refinement is concentrated near the liquid--gas interface, the embedded NURBS walls, and other regions with large solution gradients. The resulting dynamically evolving mesh is subsequently mapped to the learned density proxy used by our model.

\subsection{Dataset}
\label{sec:dataset}

Each nozzle is defined by a pair of NURBS boundary curves on the nondimensional axial domain $x \in [0, 0.7]$, parameterized by $\theta = (p_1, p_2, w)$ controlling upper-wall curvature ($p_1 \in [0.05, 0.2]$), lower-wall curvature ($p_2 \in [-0.2, -0.05]$), and global width scaling ($w \in [0.5, 1.5]$). For each geometry, incompressible two-phase flow is simulated in \texttt{Basilisk} with adaptive octree mesh refinement, advanced in steps of $\Delta t = 0.1$ for 40 timesteps. From 1{,}000 sampled geometries, we retain 797 with complete temporal trajectories (discarding the remainder due to solver failures or incomplete output), yielding approximately 31{,}000 geometry--flow snapshots.

To project the adaptive-mesh output onto a fixed $128 \times 128$ Eulerian grid, we apply Nadaraya--Watson kernel regression,
\begin{equation}
\hat{\varphi}_t(\mathbf{x}) =
\frac{\sum_{i=1}^{N_t} \varphi(\mathbf{x}_i^{(t)}) \, K_h(\mathbf{x} - \mathbf{x}_i^{(t)})}
{\sum_{i=1}^{N_t} K_h(\mathbf{x} - \mathbf{x}_i^{(t)})},
\label{eq:nw-regression}
\end{equation}
where $h = 2\max(\Delta x, \Delta y)$ denotes the Gaussian kernel bandwidth. From this projection, we derive a four-channel representation per snapshot: the mesh-cell-density proxy $\kappa$ (a log-normalized kernel density estimate of cell-center locations), volume fraction $f$, and in-plane velocities $u$ and $v$. We use simulation-level train/validation/test splits of 80/10/10 to prevent temporal leakage.

\subsection{Architecture}

Our pipeline consists of two stages: a jointly trained Stage-1 variational auto-encoder (VAE) and a separately trained Stage-2 field predictor (see \cref{fig:architecture}). The Stage-1 VAE is the core methodology of this work. It learns a shared latent representation that couples nozzle geometry and transient flow evolution, from which both the density proxy and the nozzle walls are reconstructed. The specific architectural choices used to instantiate this VAE in our experiments are an FNO encoder, a time-conditioned INR decoder, and a geometry head, described below.

\paragraph{FNO encoder.}
The encoder maps a density snapshot $\kappa \in \mathbb{R}^{1\times 128 \times 128}$, augmented with two normalized coordinate channels, to latent parameters $(\mu, \log\sigma^2) \in \mathbb{R}^{d_z}$. The architecture is a lifting Conv2d$(3 \!\to\! d_\text{model}, 1\!\times\!1)$, $L$ Fourier Neural Operator blocks~\cite{li2021} (each: spectral convolution retaining $n_\text{modes}$ Fourier modes per direction, pointwise bypass, GroupNorm, GELU), a projection Conv2d$(d_\text{model}\!\to\! d_\text{model}/2, 1\!\times\!1)$ with GELU, global average pooling, and linear heads producing $\mu$ and $\log\sigma^2$. We use $d_z=48$, $d_\text{model}=64$, $L=4$, $n_\text{modes}=16$. The encoder is time-blind by construction.

\paragraph{Time-conditioned INR decoder.}
The decoder maps $(z, t, x, y) \mapsto \hat{\kappa}(x,y;t)$ using a residual SIREN network~\cite{sitzmann2020} with FiLM conditioning~\cite{perez2018}. Spatial coordinates are embedded via multi-scale random Fourier features~\cite{tancik2020} at bandwidths $\sigma_s \in \{1, 2, 5, 10, 20, 50, 100\}$ with $n_\text{fourier}=64$ features per scale (total input dimension $= 2 \times 64 \times 7 = 896$). Time is encoded through sinusoidal features at frequencies $\omega \in \{1, 2, 5, 10, 20, 50, 100\}$. The latent code and time features are merged via a conditioning MLP $c = \text{MLP}([z; \gamma(t)]) \in \mathbb{R}^{128}$ that modulates every SIREN layer through FiLM. The decoder has hidden dimension 128 and 4 residual SIREN layers; the output is Softplus-activated for non-negativity.

\paragraph{Geometry head.}
As part of the Stage-1 VAE, a 1D INR maps $\mu$ to the nozzle walls $(y_\text{upper}(x), y_\text{lower}(x))$ at any query $x \in [0, 0.7]$: 1D Fourier features ($n=32$, $\sigma=5$) $\to$ 3-layer SIREN (hidden dimension 64) with FiLM conditioning from $\mu$ $\to$ linear 2D output.

\paragraph{Stage-2 field predictor.}
Stage-2 is a separate downstream model, trained after the Stage-1 VAE, that maps the Stage-1 reconstructed density proxy $\hat{\kappa}$ to the remaining flow variables $[\hat{f}, \hat{u}, \hat{v}] \in \mathbb{R}^{3 \times 128 \times 128}$. We implement this predictor as a U-Net~\cite{ronneberger2015}. The encoder has four downsampling stages ($1 \to 32 \to 64 \to 128 \to 256$), each with two Conv $3\times 3$ + BatchNorm + GELU blocks followed by max pooling; the decoder is symmetric with skip connections and bilinear upsampling. Output activations are sigmoid for volume fraction and identity for velocities. The Stage-2 U-Net is trained separately from Stage-1 using a weighted MSE loss, with velocity channels weighted $2\times$.

The two-stage decomposition mirrors the structure of the underlying CFD solver. Adaptive mesh refinement concentrates resolution where an error estimator identifies localized spatial complexity (interfaces, shear layers, and ligaments), producing a sparse density field that encodes where fine-scale structure exists, while velocity and volume-fraction fields are defined globally with smoother spatial variation. This suggests a natural separation: density as a spatially sparse signal, and velocity/volume fraction as globally defined fields conditioned on it. 

\subsection{Inference}
\label{sec:inference}

During inference with the trained Stage-1 VAE, we obtain the latent vector $z$ in two ways: (i) when a density snapshot is available, we set $z=\mu(\kappa_t)$ using the encoder; (ii) for a target nozzle geometry $g^\ast$ with no associated snapshot, we solve an inverse optimization problem in the latent space:
\begin{equation}
\begin{aligned}
z^\star  \;=\; & \arg\min_{\,\|z\|\le z_{\max}} \;
 \bigl\| G_\omega(z) - g^\star \bigr\|_2^2 \\
& + \lambda_\mathrm{inf} \,\bigl\| z - z_{\mathrm{anchor}} \bigr\|_2^2,
\end{aligned}
\label{eq:argmin}
\end{equation}
where $G_\omega(z)$ is the geometry decoded from the latent code by the trained geometry head with parameters $\omega$, $\lambda_\mathrm{inf} = 10^{-3}$, and $z_{\mathrm{anchor}}$ binds the optimizer to a plausible region of latent space. We use $K=8$ restarts, each anchored at one of the eight training simulations whose boundary curves are closest to $g^\ast$ in $L_2$, and retain the run with the smallest objective value. Each run is optimized with Adam for 300 steps with learning rate $5 \times 10^{-2}$ and cosine annealing. The full trajectory is then decoded by sweeping $t$ through $D(z^\ast,\cdot)$ at any desired times.

\subsection{Training and Error Metrics}

The core Stage-1 VAE is trained end-to-end by minimizing
\begin{equation}
\begin{aligned}
\mathcal{L}
= & \; 
\|\hat{\kappa}_t - \kappa_t\|_2^2
+ \lambda_g\, \|G_\omega(\mu) - g_\theta\|_2^2 \\ 
& + \beta\, \widetilde{D}_{\mathrm{KL}}\!\left(q_\phi(z \mid \kappa_t)\,\|\,\mathcal{N}(0,I)\right), 
\end{aligned}
\label{eq:loss}
\end{equation}

where $\lambda_g = 5$ weights the geometry-reconstruction term. Following the free-bits formulation of~\cite{kingma2017}, the Kullback-Leibler (KL) penalty is computed per latent dimension, averaged over the batch, and lower-bounded by a threshold $\lambda_{\mathrm{fb}} = 0.5$ before summation:
\begin{equation}
\widetilde{D}_{\mathrm{KL}}
=
\sum_{j=1}^{d_z}
\max\!\left(
\mathbb{E}_{\mathcal{B}}\!\left[D_{\mathrm{KL}}^{(j)}\right],
\lambda_{\mathrm{fb}}
\right), 
\end{equation}
where $\mathcal{B}$ denotes the minibatch, and $D_{\mathrm{KL}}^{(j)}$ the contribution of the $j$th latent dimension to the KL divergence.
This discourages posterior collapse by preventing individual latent dimensions from being regularized to near-zero information too early in training. The KL weight $\beta$ follows a linear warm-up over 100 epochs to $\beta_\text{max} = 5\times 10^{-4}$, after which a PID controller targets KL $= 18$ nats. We use AdamW ($\eta = 10^{-4}$, weight decay $10^{-4}$) with cosine annealing and gradient clipping at 1.0, batch size 16, trained for 300 epochs on a single NVIDIA A100 GPU. The Stage-2 U-Net is trained separately (Apple M4, PyTorch MPS backend) with AdamW ($\eta = 10^{-3}$), cosine annealing, batch size 32, for 20 epochs.

We report mean squared error and peak signal-to-noise ratio (PSNR $= -10\log_{10}\text{MSE}$) for density reconstruction, mean absolute error in physical coordinates for geometry, and per-channel MSE for Stage-2 field predictions. 

The combined model contains $\sim$8.8M trainable parameters (VAE: $\sim$8.7M; geometry head: $\sim$30K), while the Stage-2 U-Net contains $\sim$7M parameters.

\section{Results}
\label{sec:results}
We evaluate the full pipeline on 81 held-out test simulations (approximately 3{,}155 snapshots) that share no simulation identity with the training data (\cref{sec:dataset}). Our experiments are designed to answer four questions: (i) whether the Stage-1 latent reconstructs the AMR density proxy and nozzle geometry on unseen simulations, (ii) whether the reconstructed density field contains sufficient information to recover the remaining flow variables, (iii) whether the surrogate learns non-trivial geometry-to-flow structure beyond nearest-neighbor shape retrieval, and (iv) whether a single latent code supports efficient reconstruction of full trajectories across time, so that the same compact representation enables both geometric reconstruction and time-dependent flow decoding, including recovery of a usable latent from geometry alone via inversion of the geometry head. 

\subsection{Stage-wise Reconstruction}
\label{sec:stage-wise}

We begin with Questions (i)--(iii). \cref{tab:main_results} summarizes the main quantitative metrics across the train, validation, and test splits.

\begin{table}[tb]
\centering
\small
\caption{Main results on simulation-level train/val/test splits (637/79/81 simulations). Density and Stage-2 U-Net per-channel MSE; density PSNR; geometry MAE in physical coordinates. NN baseline column (NN): test MSE obtained by replacing the surrogate prediction with the corresponding snapshot of the nearest-neighbour training simulation in geometry space (per-point RMS distance on boundary curves, median 0.013 across the test set).}
\label{tab:main_results}
\begin{tabular}{lccccc}
\toprule
Metric & & Train & Val & Test & NN \\
\midrule
Density MSE ($\times\!10^{-3}$) & $\downarrow$      & 1.5 & 2.3 & 2.7 & 5.2 \\
PSNR (dB) & $\uparrow$                           & 31.1 & 30.2 & 30.2 & 22.9 \\
Geometry MAE ($\times\!10^{-2}$) & $\downarrow$     & 1.1 & 1.2 & 1.1 & --   \\
Stage-2 MSE, $f$ ($\times\!10^{-3}$) & $\downarrow$ & 0.7 & 1.2 & 1.3 & 2.6 \\
Stage-2 MSE, $u$ ($\times\!10^{-3}$) & $\downarrow$  & 0.9 & 1.7 & 2.5 & 5.7 \\
Stage-2 MSE, $v$ ($\times\!10^{-3}$) & $\downarrow$ & 0.7 & 1.2 & 1.6 & 3.2 \\
\bottomrule
\end{tabular}
\end{table}

\begin{figure*}[tb]
\centering
\includegraphics[width=\linewidth]{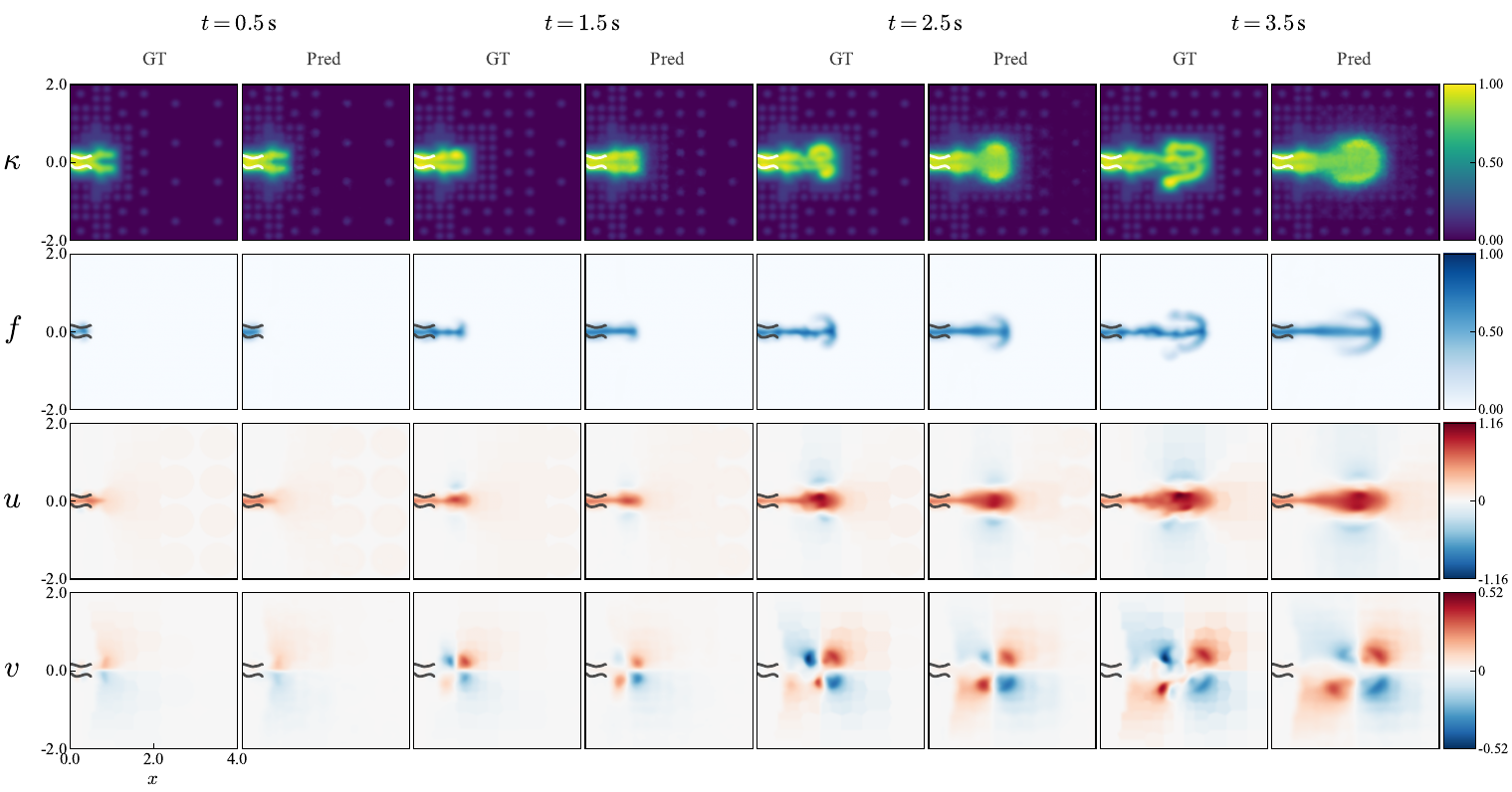}
\caption{\textbf{Full-pipeline temporal reconstruction on a held-out simulation (sim 474).} Each row shows a flow channel (density $\kappa$, volume fraction $f$, horizontal velocity $u$, vertical velocity $v$); columns show ground truth (GT) and model prediction (Pred) at four times. The latent $z$ is obtained by encoding a single \emph{late-time snapshot} ($t_{\text{max}} = 3.9$) and is held fixed for the entire trajectory. Density is reconstructed by the Stage-1 decoder from this time-independent latent, while volume fraction and velocities are predicted by the Stage-2 U-Net from the reconstructed density $\hat{\kappa}$. Predicted nozzle walls (white/gray), decoded from the same latent, are overlaid on every panel, confirming geometric consistency and showing reconstruction of both later and earlier times from a late-time encoding.}
\label{fig:recon_grid_t3.9}
\end{figure*}

\paragraph{Density and geometry reconstruction.}
The model reconstructs the transient density field with a test MSE of $2.74 \times 10^{-3}$ (PSNR 30.2~dB). Training values are $1.53 \times 10^{-3}$ (PSNR 31.1~dB), corresponding to a test/train gap of 1.8$\times$ on the reconstruction loss. The same latent $\mu$ that reconstructs the density field also decodes the upper and lower nozzle walls as functions of $x$, with a test MAE of $1.13 \times 10^{-2}$ in physical coordinates. Because the geometry head is a 1D INR, the walls can be queried at arbitrary resolution without retraining. \cref{fig:recon_grid_t3.9} overlays the predicted walls on the flow panels as a visual check.

Across reconstructed test trajectories, the integrated mesh-cell density is preserved to within $5.0\% \pm 6.2\%$ of the GT and streamwise momentum to within $11.9\%$ on average, indicating that the decoded fields respect the basic conservation structure of the underlying flow without explicit conservation losses being imposed during training.

\paragraph{Stage-two flow field prediction.}
The Stage-2 U-Net maps the reconstructed density $\hat{\kappa}$ to volume fraction and velocity components, achieving per-channel test MSE of $1.33 \times 10^{-3}$ ($f$), $2.50 \times 10^{-3}$ ($u$), and $1.60 \times 10^{-3}$ ($v$), corresponding to test R$^2$ of 0.69, 0.78, and 0.61 respectively. Since the predictor operates only on the reconstructed AMR-density proxy rather than the full simulator state, these results show that the proxy captures substantial information about the spatial organization of the remaining flow variables and provides an effective conditioning signal for downstream reconstruction, despite having to generalize across held-out nozzle geometries and transient breakup regimes. The strongest performance is observed for the streamwise velocity $u$, consistent with the dominant axial structure of the jet, whereas reconstruction of the weaker cross-stream component $v$ is inherently more demanding.

\paragraph{Comparison to geometry-space nearest-neighbor.}
To check that the surrogate has learned non-trivial geometry-to-flow structure rather than performing implicit shape retrieval, we compare against a nearest-neighbor(NN)-in-geometry baseline: for each test snapshot we substitute the corresponding snapshot of the training simulation whose boundary curves are closest in L2 distance, and report the resulting MSE in \cref{tab:main_results}. The surrogate beats the baseline by $\approx 2\times$ on mean MSE across all four channels, despite a tightly bounded test-set geometry-distance distribution (per-point RMS in $[0.009, 0.020]$). The baseline's MSE is heavy-tailed (per-channel std/median ratio of 2--4), reflecting test geometries on which small wall changes drive large flow changes (the nonlinear geometry-flow sensitivity flagged in \cref{sec:introduction}). The surrogate handles these sensitive cases without the same degradation, indicating that the latent encodes flow structure beyond what is recoverable by retrieving the nearest training shape.

\subsection{Temporal Reconstruction}
\label{sec:temporal}

We next address Question (iv) for inference path (i) (\cref{sec:inference}), i.e., latent inference from a density snapshot via the encoder. At inference, full 39-timestep trajectory reconstruction, including Stage-2 field prediction and geometry decoding, takes $0.045$ seconds per simulation on average (20 test cases; Apple M4 GPU), yielding a $>10^{4}$ speed-up over Basilisk CFD (45--90 minutes on 12 CPU cores).

\cref{fig:recon_grid_t3.9} shows a representative test simulation at four times spanning the simulation window, reconstructed through the full pipeline: density $\hat{\kappa}$ from the Stage-1 decoder, volume fraction $\hat{f}$ and velocities $\hat{u}, \hat{v}$ from the Stage-2 U-Net, and nozzle walls from the geometry head. These are decoded from a single latent $\mu$ encoded from a late-time GT snapshot ($t_{\text{max}} = 3.9$) of the trajectory. The full trajectory is generated by sweeping $t$ through the decoder without re-encoding, including reconstruction of earlier states (e.g. $t = 0.5, 1.5, 2.5, 3.5$), demonstrating that the latent captures global trajectory information rather than purely forward dynamics. For completeness, reconstructions obtained from an early-time GT encoding ($t_{\text{min}} = 0.1$) are provided in \cref{app:early_time_encoding}. The decoder captures large-scale jet development and shear-layer roll-up, but fails to resolve the formation and subsequent fragmentation of thin ligaments, instead producing overly smooth, diffused structures at later times.

\begin{figure}[tb]
\centering
\includegraphics[width=\linewidth]{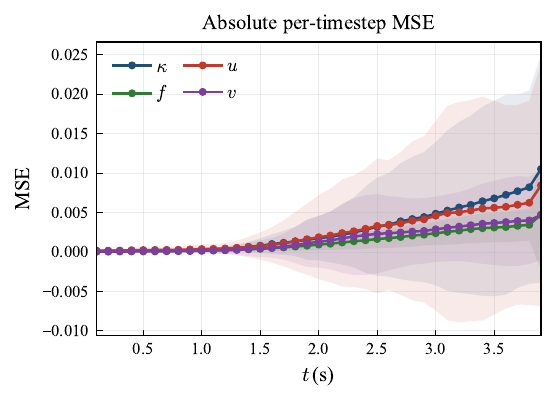}
\caption{\textbf{Per-timestep reconstruction error across the 81 test 
simulations.} Absolute per-channel MSE; shaded bands show 
$\pm$ one standard deviation across test simulations.}
\label{fig:mse_vs_time}
\end{figure}

\paragraph{Per-timestep error.}
\cref{fig:mse_vs_time} reports per-timestep reconstruction MSE across the 81 test simulations for all four channels. The density error increases gradually toward later times, with a corresponding rise in the errors of $f$, $u$, and $v$ due to the Stage-2 dependence on $\hat{\kappa}$. This behavior is consistent with increasing intrinsic difficulty of the flow: as the jet evolves, interfaces thin, curvature increases, and density gradients become sharper and more spatially localized. These features occupy a progressively smaller fraction of the domain while carrying higher-frequency content, making them harder to represent at fixed latent and decoder capacity. The Stage-2 fields inherit this trend because they are conditioned on $\hat{\kappa}$, so errors in fine interfacial structure propagate into velocity and volume-fraction predictions. This trend is visible across many held-out simulations, but a small high-error tail emerges at later times. 

The failure case in \cref{app:failure_case} (\cref{fig:failure_case}), ranked among the worst 5\% by composite reconstruction error at $t = 3.9$s, exposes the dominant failure mode: thin ligaments and sharply localized breakup structures are smoothed (loss of high-frequency interfacial detail), while the macroscopic jet trajectory exhibits increasing deviation at later times.

\begin{table}[tb]
\centering
\small
\caption{Geometry-targeted inversion on 81 held-out test simulations. \emph{Oracle}: $z=\mu(\kappa_s(t_{\max}))$ from the ground-truth density snapshot. \emph{Argmin}: $z^\ast$ from \cref{eq:argmin}, anchored at the $K=8$ training simulations geometrically closest to $g^\ast$. \emph{NN}: latent taken from the training simulation with nearest boundary curves in $L_2$. Density MSE is reported as both median and mean because the mean is sensitive to a small number of high-error cases.}
\label{tab:geom_inversion}
\begin{tabular}{lcccc}
\toprule
Metric & & Oracle & Argmin & NN \\
\midrule
Geometry MAE ($\times\!10^{-3}$) & $\downarrow$ & 10.2 & 0.4 & 14.2 \\
Density MSE median ($\times\!10^{-3}$) & $\downarrow$ & 2.3 & 2.5 & 2.7 \\
Density MSE mean ($\times\!10^{-3}$) & $\downarrow$ & 3.7 & 4.0 & 4.4 \\
PSNR (dB) & $\uparrow$ & 28.3 & 28.7 & 28.4 \\
Stage-2 MSE, $f$ ($\times\!10^{-3}$) & $\downarrow$ & 2.7 & 2.7 & 2.8 \\
Stage-2 MSE, $u$ ($\times\!10^{-3}$) & $\downarrow$ & 5.2 & 5.5 & 6.1 \\
Stage-2 MSE, $v$ ($\times\!10^{-3}$) & $\downarrow$ & 2.5 & 2.6 & 2.7 \\
\bottomrule
\end{tabular}
\end{table}

\subsection{Geometry-targeted Latent Inversion}
\label{sec:geom_inversion}

Regarding inference path (ii) (\cref{sec:inference}), we assess whether the same latent space remains usable for full-trajectory reconstruction when the latent is recovered from geometry alone rather than from an observed density snapshot. We test this geometry-only path on the 81 held-out test simulations. For each test geometry $g_s$, we recover $z^\ast_s$ by \cref{eq:argmin} and decode the entire trajectory; the encoder is never given a snapshot of simulation $s$. \cref{tab:geom_inversion} compares this inference path to two baselines: an \emph{encoder oracle} that uses $z = \mu(\kappa_s(t_{\max}))$ from the ground-truth density (an upper bound on what the latent supports), and a \emph{nearest-neighbor-in-geometry} baseline that substitutes the training simulation with the closest boundary curves in $L_2$.

The recovered latent $z^{\ast}$ decodes density to a median $\kappa$ MSE between the encoder oracle (which sees the ground-truth density snapshot) and a nearest-neighbor-in-geometry baseline, and outperforms the latter on 47/81 simulations, compared with 50/81 for the oracle. Geometry alone is therefore sufficient to recover a usable latent representation in practice: gradient-based latent optimization closes 55\% of the gap to encoder-based inference without ever observing $\kappa$. The same ordering is observed for the Stage-2 channels, with gradient-based latent optimization consistently closer to the oracle than nearest-neighbor retrieval on $f$, $u$, and $v$ (\cref{tab:geom_inversion}), indicating that inversion preserves downstream flow-state quality rather than only improving geometry or density reconstruction. The worst-case test simulation (sim 790, $\kappa$ MSE $= 0.032$ versus a typical value of $\sim\!0.003$) lies near the edge of the geometric distribution ($d_\mathrm{NN} = 0.25$, well above the median of $\sim\!0.19$); on this case, both the encoder oracle ($0.040$) and nearest-neighbor baseline ($0.056$) fail more severely than gradient-based latent optimization, indicating a Stage-1 representational limit on out-of-distribution geometries rather than a failure of inversion itself. We do not claim that gradient-based refinement strictly dominates geometry retrieval at this sample size; rather, its value is that it provides a tractable inference pathway from geometry alone when no density snapshot is available at inference time.

\section{Discussion}
\label{sec:discussion}

\paragraph{Summary.}
We present a geometry-conditioned latent surrogate for transient two-phase nozzle flow, trained on 797 simulations with a clean simulation-level train/validation/test split. The central result is that a single shared latent representation can couple nozzle geometry and transient flow evolution when learned from the AMR cell-density field rather than the full simulator state. On held-out simulations, this is sufficient to reconstruct transient density evolution with PSNR $\approx 30.2$~dB, recover nozzle geometry with MAE $\approx 1.0 \times 10^{-2}$, and, through a lightweight Stage-2 predictor, recover velocity and volume-fraction fields from the reconstructed density. The results therefore support a broader methodological claim: in adaptive multiphase CFD, the solver’s own refinement structure can serve as a compact and learnable representation of flow-relevant dynamics. The geometry-targeted inversion results further suggest that this shared latent remains usable even when only target geometry is available at inference time, making the surrogate a plausible building block for future geometry-driven design workflows. 

\paragraph{Limitations.}
Several limitations constrain the scope of our claims. Training uses 797 simulations drawn from a three-parameter NURBS nozzle family at a single Reynolds--Weber regime; extension to higher-dimensional design spaces, broader operating conditions, and more diverse breakup regimes remains open. Although the simulation-level split prevents snapshot-level leakage, the test set contains only 81 simulations, so the reported metrics should be interpreted with that sample size in mind. The two-stage decomposition means that the Stage-2 predictor is trained independently from the latent model, so reconstruction errors in density propagate into velocity and volume-fraction prediction. In addition, the learned representation is tied to the AMR structure produced by the particular solver and refinement strategy used here, and its transferability across solvers or mesh-adaptation criteria has not yet been established. Finally, all results are evaluated within the learned surrogate itself; validation against independent CFD reruns or experimental measurements would be necessary before operational use and lies beyond the scope of the paper.

\paragraph{Future work.}
Future work follows directly from the architecture. Adding a droplet-distribution head conditioned on the decoded flow fields would complete the causal chain from geometry to atomization-relevant observables, such as droplet-size distribution and satellite counts, constrained primarily by the availability of labeled breakup data. The learned latent also opens the door to generative inverse design
as a natural next step. We plan to validate optimized designs through independent Basilisk simulations, closing the loop between learned design and high-fidelity CFD. More broadly, we view this work as a step toward surrogate models that exploit, rather than discard, the structure induced by adaptive scientific solvers.

\section*{Acknowledgements}
The authors acknowledge computational resources and support provided by the Imperial College Research Computing Service (\url{http://doi.org/10.14469/hpc/2232}). J.H.~Ramlau acknowledges support from the CMAC Engineering and Physical Sciences Research Council (EPSRC) Centre for Doctoral Training in Cyber-Physical Systems for Medicines Development and Manufacturing (CEDAR) (Grant Ref. EP/Y035593/1). F.~Hastedt acknowledges support from EPSRC grant EP/S023232/1. T.~Birdal acknowledges support from the UKRI Engineering and Physical Sciences Research Council (EPSRC) through a Future Leaders Fellowship (grant number MR/Y018818/1). N.S.~Basha acknowledges support from the EPSRC Impact Acceleration Account (IAA) grant.


\section*{Impact Statement}

Efficient surrogate models for atomization processes can substantially accelerate the computational design of nozzles in applications such as drug delivery, fuel injection, and industrial coating. By reducing reliance on expensive high-fidelity simulations, such methods enable broader exploration of design spaces and may facilitate more rapid and cost-effective engineering iteration.

The potential risks associated with this work are limited. The study focuses on forward modeling within a well-established physical and engineering domain and does not involve human subjects, personal data, or systems that directly inform automated decisions affecting individuals. No sensitive or dual-use data is employed, and the models are intended for scientific and engineering analysis rather than deployment in safety-critical or human-facing decision systems.

\section*{Ethics Statement}

This work adheres to the principles outlined in the ICML code of research ethics, including transparency, research integrity, and responsible dissemination of results. All datasets used are derived from established numerical simulation pipelines. We have considered the environmental impact of training and experimentation, and computational resources were used judiciously, with large-scale training runs performed only when necessary to support the conclusions of the study.

\section*{Reproducibility Statement}

The source code for training and inference is publicly available at \url{https://github.com/Computational-Design-Intelligence/spray-surrogate}. While the code is released openly, the full datasets and trained model checkpoints are not publicly distributed at this time due to storage and governance constraints; however, they are available from the authors upon reasonable request.

All experimental details, including model architectures, training procedures, and evaluation protocols, are described in sufficient detail in the main text to enable replication. Where applicable, hyperparameters are specified to facilitate reproducibility of results.

\bibliographystyle{icml2026}
\bibliography{bibliography_icml}

\newpage
\appendix
\onecolumn

\section{Additional Reconstructions}
\label{app:additional_recons}

\subsection{Early-Time Encoding}
\label{app:early_time_encoding}

\begin{figure}[H]
\centering
\includegraphics[width=\linewidth]{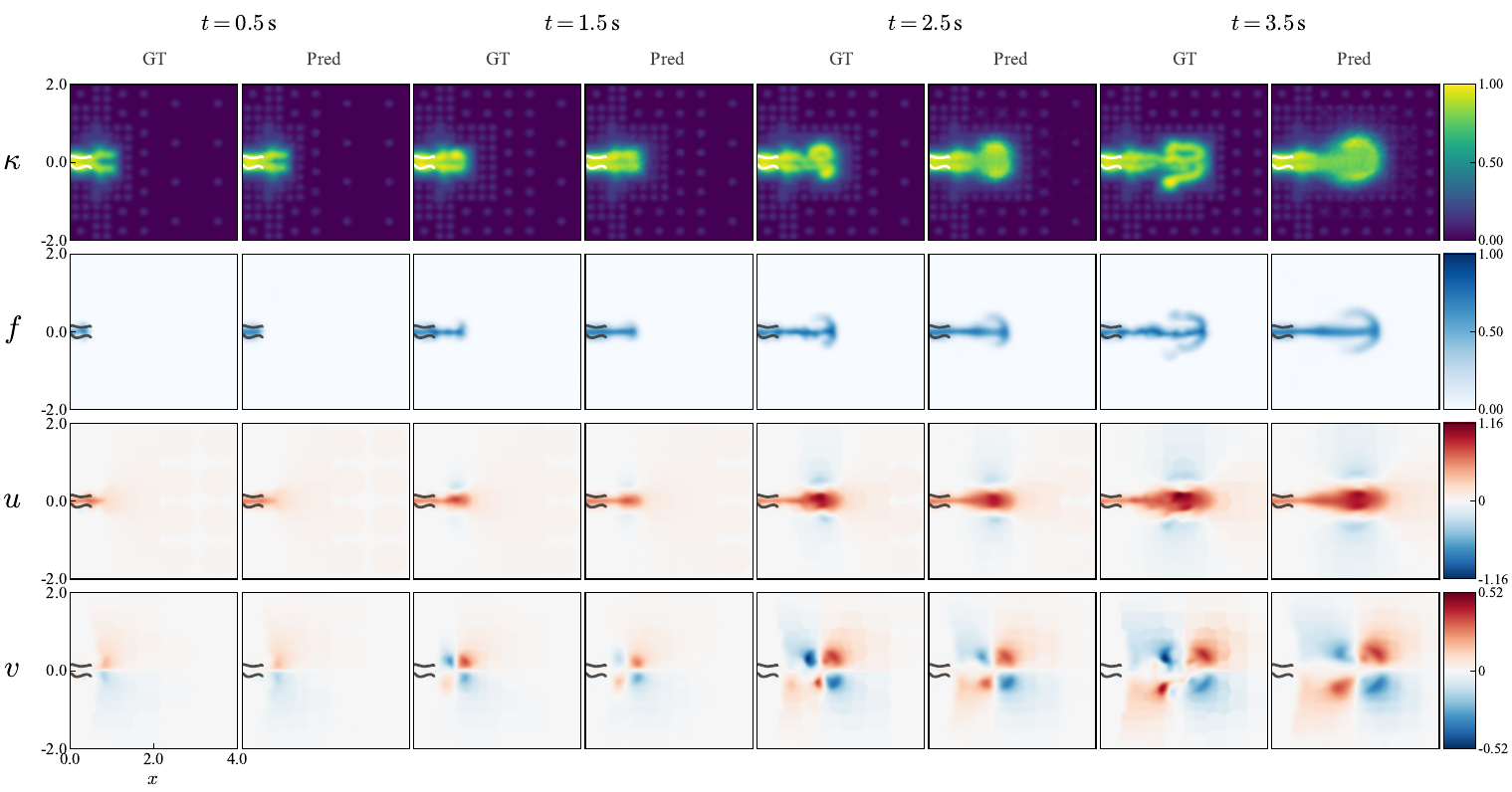}
\caption{\textbf{Effect of early-time encoding on full-pipeline reconstruction (sim 474).} Same setup as Fig.~\ref{fig:recon_grid_t3.9}, but the latent $z$ is obtained from an \emph{early-time snapshot} ($t_{\text{min}} = 0.1$) and held fixed for the full trajectory reconstruction. All channels (density $\kappa$, volume fraction $f$, velocities $u,v$) and geometry are decoded as in the main figure, showing qualitatively similar reconstruction behavior.}
\label{fig:recon_grid_t0.1}
\end{figure}

\newpage 
\subsection{Representative Failure Case}
\label{app:failure_case}
\begin{figure}[H]
\centering
\includegraphics[width=\linewidth]{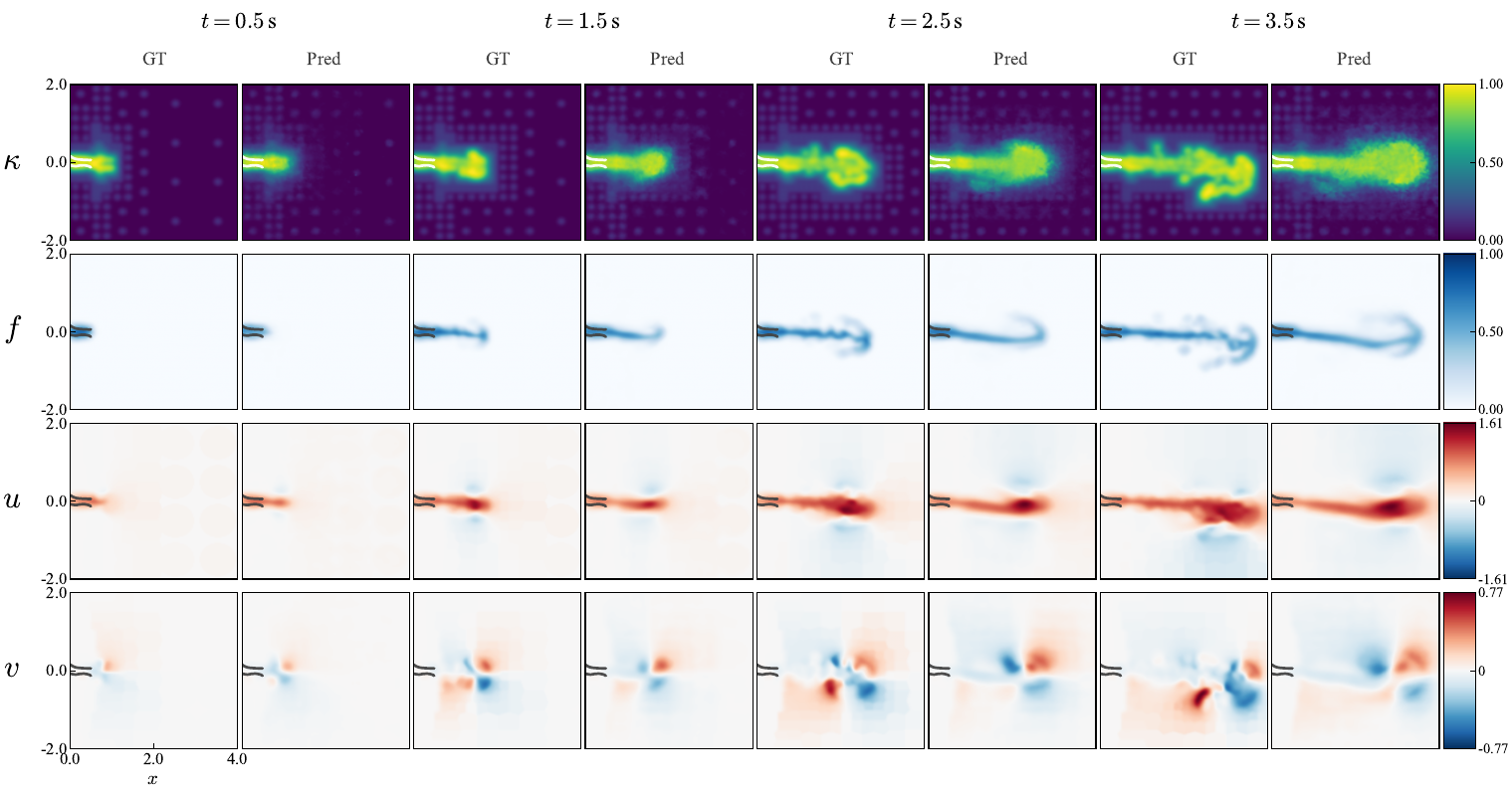}
\caption{\textbf{Representative failure case on a held-out test simulation (sim 366).} This example ranks among the worst 5\% of held-out simulations by composite reconstruction error at $t=3.9$. The model smooths thin ligaments and under-resolves sharply localized late-time breakup structures, with visible errors also appearing in the macroscopic jet trajectory.}
\label{fig:failure_case}
\end{figure}


\end{document}